\documentclass[aps,prb,twocolumn,showpacs,amsmath,amssymb]{revtex4}
\begin{document}

\title{\bf Negative heat capacity of sodium clusters}

\author{
 Juan A. Reyes-Nava,
 Ignacio L. Garz\'on, and
 Karo Michaelian
}

\affiliation{
    Instituto de F\'{\i}sica,
    Universidad Nacional Aut\'onoma de M\'exico,
    Apartado Postal 20-364, 01000 M\'exico D.F., M\'exico 
}

\date{\today}

\begin{abstract}
Heat capacities of Na$_{N}$, $N$ = 13, 20, 55, 135, 142, and 147, 
clusters have been investigated using a many-body Gupta potential and
microcanonical molecular dynamics simulations.
Negative heat capacities around the cluster melting-like transition have
been obtained for $N$ = 135, 142, and 147, but the smaller
clusters ($N$ = 13, 20, and 55) do not show this peculiarity.
By performing a survey of the cluster potential energy
landscape (PEL), it is found that the width of the distribution function
of the kinetic energy and the spread of the
distribution of potential energy minima (isomers), are
useful features to  determine the different behavior of the 
heat capacity as a function of the cluster size.
The effect of the range of the interatomic forces 
is studied by comparing the heat capacities of the Na$_{55}$
and Cd$_{55}$ clusters. It is shown that by decreasing the
range of the many-body interaction, the distribution of isomers
characterizing the PEL is modified appropriately to
generate a negative heat capacity in the Cd$_{55}$ cluster.
\end{abstract}


\maketitle

\section{Introduction}
Negative microcanonical heat capacity in 
atomic and molecular clusters was theoretically
predicted by considering simple models of the distribution
of local minima that characterize the potential energy 
landscape (PEL) of clusters \cite{Bixon}.
In that study, it was found that for high values of the parameter
involving the ratios of the vibrational frequencies corresponding
to the global and local isomers, the caloric curve displays
an S-shaped loop, with a negative heat capacity in the
vicinity of the melting point \cite{Bixon}.
In another study on the solid-liquid transition of clusters \cite{Labas},
it was shown that in microcanonical simulations of 
Lennard-Jones clusters, an increase in total energy causes a 
temperature reduction. This effect was related to the broadening
of the cluster kinetic-energy distribution toward lower energy 
values \cite{Labas}.

Although the existence of negative heat capacity in 
physical systems like stars or
star clusters \cite{Thirring,Bell}, and in 
fragmenting nuclei \cite{Gross,Agostino} is well-known,
this peculiar effect gained a lot of interest in the field of atomic
and molecular clusters due to recent experimental results
where a negative heat capacity was measured for a 147-atom
sodium cluster \cite{Schmidt}. In this study, the photofragmentation
mass spectra was used to measure the internal energy of free, mass  
selected clusters with known temperature. These measurements were
used to determine 
the microcanonical caloric curve of the Na$^{+}$$_{147}$ that
shows the characteristic S-shaped (backbending) feature,
indicating a negative heat capacity \cite{Schmidt}.
The negative value of the microcanonical heat capacity was
interpreted by considering that a finite system upon melting tries to avoid
partly molten states and prefers to convert some of its
kinetic energy into potential energy \cite{Schmidt,Schmidt1}.
This peculiarity has been attributed to the non-additivity
of the total energy of a cluster with finite size \cite{Schmidt,Schmidt1}.

Microcanonical heat capacities of metal clusters have been 
theoretically investigated using constant-energy molecular
dynamics (MD) with many-body potentials 
\cite{Gallego,Calvo,Jell} and an orbital-free version
of the first-principles MD method
\cite{Agua1,Agua2}. In these studies, heat capacities 
of fcc transition and noble metal clusters with up to 
23 atoms were calculated to characterize their melting-like
transition \cite{Gallego}. In another study, on the melting of sodium
clusters \cite{Calvo}, the microcanonical caloric curve
of the Na$_{55}$ cluster was obtained. However, in not one of these studies
was a signature of a negative heat capacity found.
Similar results, indicating the non-existence of a negative heat
capacity in constant-energy orbital-free first-principles MD simulations
of larger sodium clusters (Na$_{55}$, Na$_{92}$, and Na$_{142}$),
were obtained \cite{Agua2}. 
Nevertheless, in such calculations the simulation time
employed was too short to obtain converged results \cite{Agua2}.
On the other hand, in microcanonical MD simulations of Al$_{N}$,
$N$ = 7, 13, 55, and 147, clusters, a negative heat capacity was
obtained for the larger Al$_{147}$ cluster \cite{Jell}.

In the present work, motivated by the availability of
experimental techniques that allow the measurement of the
microcanonical heat capacity and other thermal properties
of mass selected metal clusters \cite{Haber1,Haber2,Haber3}, 
we theoretically investigate the behavior of the heat capacity
of sodium clusters in the size range of 13-147 atoms.
In our approach, constant-energy MD simulations are performed
using a phenomenological many-body potential
that mimics the metallic bonding of sodium clusters.
This approximation allows us to use simulation times of the
order of $\sim$ 50 ns, in order to obtain converged
averages of the microcanonical heat capacity
and other cluster thermal properties.
Our main objective is to gain additional insights into the
conditions that determine if a cluster has a negative
heat capacity. The main finding of this work shows that
the width of the distribution function of the kinetic energy 
and the spread of the distribution of the potential energy minima
(isomers), characterizing the PEL,
are useful
features to determine the signature of the
cluster heat capacity.
In section II, we provide the theoretical background on which
this study is based. The results and their discussion are
given in section III, and section IV contains a summary and the
conclusions of this work.

\section{Theoretical background}
The heat capacity and temperature of sodium
clusters as a function of the cluster total energy are calculated
through constant-energy MD simulations using the microcanonical
expressions derived in Refs. \onlinecite{Pearson,Sugano}:
\begin{equation}
\label{C}
\frac{C}{Nk_{B}}= \left[ N-N(1-\frac{2}{3N-6}) \langle K \rangle
\langle K^{-1} \rangle \right]^{-1}
\end{equation}

\begin{equation}
\label{tem}
T=\frac {2\langle K \rangle }{(3N-6)k_{B}},
\end{equation}
where $K$ is the
kinetic energy of the cluster, $k_{B}$ is the Boltzmann constant,
and $\langle$...$\rangle$ denotes a time average.
In these formulas, 3$N$ was changed to 3$N$-6, the
number of degrees of freedom of the system, since the calculations
are performed for a non-translating and non-rotating cluster
in a three-dimensional space
(the position of the center of mass was fixed and the 
total momentum was held to zero during the simulations).

In our implementation of the constant-energy MD method, 
the Newton's equations of motion are solved with the Verlet
algorithm \cite{Verlet} using a time step of 2.4 fs, which
provides total energy conservation within 0.001 $\%$. A typical
calculation consists in heating up a cluster from its lowest-energy
solid-like configuration until it transforms into a liquid-like
cluster. To simulate this procedure the cluster total energy
is increased in a step-like manner by scaling up the atomic
velocities. For each initial condition the cluster was equilibrated
during 10$^4$ time steps and the time averages of the physical
quantities are calculated using 10$^7$ time steps. This averaging
time is increased by a factor of 2 when the cluster is in the
region of the solid-to-liquid transition in order to ensure
the calculation of fully converged averages.

To model the metallic bonding in sodium clusters
we used the many-body Gupta potential \cite{Gupta}, 
which is based on the second
moment approximation of a tight-binding hamiltonian \cite{Cleri}.
Its analytical expression is given by:
\begin{equation}
\label{V}
V = \sum_{i=1}^{N} V_i
\end{equation}
\begin{equation}
V_i =  A \sum_{j \ne i} e^{-p \left(\frac{r_{ij}}{r_0}-1 \right)} -
  \xi \left( \sum_{j \ne i}
  e^{-2q \left(\frac{r_{ij}}{r_0}-1 \right)} \right)^{\frac{1}{2}}
\label{Gupta}
\end{equation}
where $r_0$, $A$,
$\xi$, $p$, and $q$ are adjustable parameters 
\cite{Cleri}. For sodium clusters these parameters have been
fitted to band structure calculations \cite{Li}. Their values
are: $A$=0.01595 eV, $\xi$=0.29113 eV, $r_0$=6.99 bohr, $p$=10.13,
and $q$=1.30 \cite{Li}.
This phenomelogical many-body potential has been used to study
the melting-like transition in sodium clusters of different sizes
using Monte Carlo (MC) \cite{Calvo}
and constant-energy
MD simulations \cite{Reyes}.
A good qualitative agreement has been obtained between  
structural and thermal properties calculated using the Gupta
potential \cite{Calvo,Reyes} and those generated from
first-principles methods \cite{Agua1,Agua2}. 
An additional advantage in using
this potential is that it allows 
simulation times of the order of 50 ns, 
necessary to obtain fully converged
time averages in the melting-like transition region.

\section{Results and Discussion}
The microcanonical heat capacities of the Na$_N$, $N$ = 13, 20,
55, 135, 142, and 147, clusters, calculated using Eq. (1),
are displayed in Fig. \ref{Fig1}. For $N$ = 13, 20, and 55, they are
continuous functions of the cluster total energy showing
a maximum value that is characteristic of a melting-like
transition
\cite{Calvo,Agua1,Agua2,Reyes}.
On the other hand, the heat capacity of the larger clusters
($N$ = 135, 142, and 147) shows  two discontinuity points and
a continuous negative-valued interval between them.
This peculiar behavior in the heat capacity is related with
a backbending loop in the caloric curve (temperature
as a function of the total energy) \cite{Bixon,Labas,Wales1,Wales}.
In fact, our calculated microcanonical caloric curves 
of Na$_{135}$, Na$_{142}$, and Na$_{147}$ show the
backbending loop at the same energies where the heat
capacity takes negative values (see the caloric curves
shown in Fig. 2 of Ref. \onlinecite{Reyes}).
In previous studies, the negative slope (backbending loop) of
the microcanonical caloric curve has been 
attributed to a peculiar behavior
of the cluster entropy as a function of energy
that shows a \it dent \rm  with inverted curvature
in the region of the solid-liquid 
transition \cite{Bixon,Labas,Schmidt,Schmidt1,Wales}.

In the present work, we analyze the behavior of the microcanonical
heat capacity of sodium clusters from a different perspective.
First, we consider Eq. (1) and obtain the condition
to have a negative value in the heat capacity:
\begin{equation}
Z_E \, \equiv \, \langle K \rangle  \langle K^{-1} \rangle
 \, \,\,   >  \, \, \, \frac{3N-6}{3N-8}.  
\label{Z}
\end{equation}
Figure \ref{Fig2} display the values of $Z_{E}$ (black dots)
as a function of the
cluster total energy $E$ that were calculated from
a time average, using the MD trajectories. In the same scale 
the threshold
value $Z_{c} \equiv (3N-6)/(3N-8)$ for each cluster size is given.
In Fig.  \ref{Fig2} , it can be graphically seen how the
relative difference between $Z_{E}$ and $Z_{c}$ changes
with the cluster size. For the three smaller clusters
(see panels (a), (b), and (c) in Fig.  \ref{Fig2}  )
the $Z_E$ values do not overcome the threshold value,
whereas for the three larger clusters there is a range
of total energy where $Z_{E}$ satisfy the condition
to have negative heat capacities (see panels (d), (e), and
(f) in Fig. \ref{Fig2} ).

In order to investigate what determines a
negative value of the heat capacity, we consider the quantity
$Z_{E}$. This is the product of the averages
of the kinetic energy and of the inverse of this
quantity, and therefore, its value will depend on the distribution
function of the kinetic energy, $g_{E}(K)$. 
The average of any function of the
kinetic energy $f(K)$ can be obtained
through the following expression:
\begin{equation}
{\langle {f_{E}} \rangle}_g =  \int f(K) g_{E}(K) dK. 
\label{f}
\end{equation}
Since the distribution function of the kinetic energy, $g_E(K)$,
determines the behavior of $Z_{E}$, 
it is useful to analyze $g_{E}(K)$ at different
values of the cluster total energy $E$.
The calculation of this quantity is straightforward 
from the constant-energy MD simulations. 
Figure \ref{Fig3} shows $g_{E}$ as a function of the normalized
mean deviation
$\delta K$ = ($K$ - $\langle K \rangle$)/ $\langle K \rangle$
for three different energies,
corresponding to the cases where the cluster is in
the solid- and liquid-like phases, and at the middle
of the melting-like transition.
The analysis of $g_{E}$ as a function of $\delta K$, instead of
$K$, has the advantage that it allows the comparison,
on the same scale, 
of the lineshapes of this
function at different cluster energies and for different
cluster sizes.

As a general trend, it is found that $g_E$
becomes narrower for increasing cluster sizes, indicating
a larger relative dispersion of the kinetic energy values
for the smaller clusters.
This result is expected since it confirms  the increment
of fluctuations in
kinetic energy of a physical system that decrease in
size. A common characteristic of $g_E$, existing
in the six clusters investigated, is the larger broadening
of the distribution function when the cluster is at 
the melting-like
transition. At lower (solid-like phase) and 
higher (liquid-like phase) energies,
the width of $g_E$ is smaller, whereas at
the phase transition the fluctuations in kinetic 
energy, as expected, should increase.
  
For the three smaller clusters which do not have negative
heat capacity, $g_{E}(\delta K)$ shows
a nearly symmetric lineshape independent of the
cluster energy (see panels (a), (b), and (c) of Fig. \ref{Fig3} ). 
In contrast, 
Na$_{142}$ and Na$_{147}$, that show a negative heat capacity,
have a distribution function $g_{E}(\delta K)$ with a
shoulder towards positive values of $\delta K$,
at energies in the middle of the melting region (see panels
(e) and (f) of Fig. \ref{Fig3} ).
Although this difference in the distribution function of the kinetic
energy could be a useful feature
to determine the existence of a negative heat capacity, 
the Na$_{135}$ cluster would be an exception to this rule since
its $g_{E}(\delta K)$  does not show a resolved shoulder
in its lineshape (see panel (d) in Fig. \ref{Fig3}),
but it has negative heat capacity.
 
On the other hand, a characteristic of $g_{E}$ that 
would be useful to determine
the sign of the heat capacity is the width of the
distribution function, which can be obtained through its
second moment:
\begin{equation}
{\langle (\delta K)^2 \rangle}_g =  
\int (\delta K)^2 g_{E}(\delta K) d(\delta K).
\label{sigma}
\end{equation}
The second moment of $g_{E}$ 
corresponds to the second term in the
expansion of $Z_E$,  which can be obtained from the
left hand term of Eq. (5) \cite{Pearson,Sugano}:
\begin{equation}
Z_E =  1 + {\langle (\delta K)^2 \rangle}_g + ....
\label{Zbar}
\end{equation}
By taking terms up to second order in 
this expansion, assuming that
$ {\langle \delta K \rangle}_g << 1 $, $Z_E$ can be approximated by:
\begin{equation}
Z_E \approx Z_{E}^{(2)} = 1 + {\langle (\delta K)^2 \rangle}_g. 
\label{Z2}
\end{equation}
Since the $Z_{E}^{(2)}$ values can be 
calculated using $g_E$ and Eq. (7), it is possible
to check the validity of this approximation, which can only
be applied to systems with a finite number of particles.
Fig. \ref{Fig2} shows the values
of $Z_{E}^{(2)}$ (stars) as a function of the cluster energy. It can be seen
that the difference with $Z_E$ is small for the three smaller clusters
and negligible for the three larger ones.
Then, $Z_{E}^{(2)}$ can be considered as a 
quantitative measure of the width
(second moment) of the distribution function of the kinetic energy,
and can be used to determine the sign of the heat capacity.
Figure \ref{Fig4} shows the maximum values of $Z_{E}^{(2)}$ 
(black dots) calculated for energies at the middle of the melting 
transition, and their comparison with the threshold values $Z_c$
as a function of the cluster size (full line). 
From this figure, it is obvious
that although the width of $g_E$ is relatively large for
the three smaller clusters, the corresponding
maximum values of $Z_{E}^{(2)}$
are below the threshold $Z_c$, and therefore these
clusters do not show a negative heat capacity. On the other hand, 
the width of $g_E$ for Na$_{135}$, Na$_{142}$, and Na$_{147}$
is smaller than for Na$_{13}$, Na$_{20}$, and Na$_{55}$,
however, $Z_c$ is a faster decreasing function
of the cluster size $N$, such that
the maximum of $Z_{E}^{(2)}$ 
lies above the threshold values, indicating that the
larger clusters have negative heat capacities.
Then, the above results suggest that the width of the distribution function
of the kinetic energy is a useful property to determine the
sign of the heat capacity of clusters. However, this
quantity not only depends on the cluster size, 
but also on the characteristics
of the PEL. 

To illustrate the importance
of the topology of the PEL we have investigated
the behavior of the heat capacity of
55-atom clusters using  the many-body Gupta potential
\cite{Gupta}, shown in Eqs. (3) and (4), for the
different metals listed in Tables I and III of
Ref. \onlinecite{Cleri}, and in Table II of
Ref. \onlinecite{Li}. Our results show that 
the Cd$_{55}$ cluster (with the following parameter values \cite{Cleri}: 
$A$=0.0416 eV, $\xi$=0.4720 eV, 
$p$=13.639,
and $q$=3.908) has a negative heat capacity, but none of
the other 55-atom clusters show this peculiarity.
The upper insets of
Fig. \ref{Fig4} show the calculated caloric curve
with the corresponding backbending loop
and the heat capacity with negative values for a range
of total energy  values of the Cd$_{55}$ cluster. The inset 
at the lower right corner of Fig. 4
shows $g_E$ as a function of the normalized 
mean deviation of the kinetic energy for both,
the Na$_{55}$ and Cd$_{55}$, clusters.
It can be
seen that the broadening of $g_E$ is larger in Cd$_{55}$ than
in Na$_{55}$, such that, it
generates a maximum value of $Z_{E}^{(2)}$ (this value
corresponds to the point represented by a star in Fig. 4)
that overcome the threshold $Z_c$,
and consequently, the Cd$_{55}$ cluster display
a negative heat capacity. This comparison with the
Na$_{55}$ cluster which does not show this peculiarity, indicates
that although both clusters have the same size, their dynamical
properties defined by their corresponding PEL's, generate different 
behavior in their heat capacities.

In order to investigate the influence of the PEL
on the different
widths of the distribution function of the kinetic energy
of the Na$_{55}$ and  Cd$_{55}$ clusters, 
further studies are
necessary. In this direction, the calculation of short-time
averages of the kinetic energy and periodic quenchings of instantaneous
configurations during the MD trajectories allow us to
obtain the distribution of potential energy minima (isomers)
that are accessible at different
cluster energies \cite{Wales1}.
Figure \ref{Fig5} display the normalized 
distribution of potential energy minima, obtained
by periodical quenchings using MD trajectories at a total
energy where the cluster is at the middle of the melting
transition, for the Na$_{55}$ and Cd$_{55}$ clusters.
It can be notice that the number of isomers with higher energy
relative to the global minimum,
are larger for the
Cd$_{55}$ cluster in comparison with the results obtained
for Na$_{55}$. This result can be explained by taking into
account that the range of the interatomic forces is shorter
in Cd than in Na clusters, mainly due to the higher value
of the $q$ parameter in the many-body Gupta potential
\cite{Karo3}. The physical reason for the larger number of
minima at short range is the loss of accessible configuration
space as the potential wells become narrower, thus
producing barriers where there are none at long range
\cite{Doye}.

To show how the distribution of potential energy minima determine the
broadening of the distribution function of the kinetic energy, we
approximate the complex topology of the PEL by a 
set of independent harmonic 
potential wells in the $3N-6$ dimensional space. Each one of
these wells is associated to the different 
potential energy minima forming the distribution 
of isomers shown in Fig. 5. For each potential energy
minimum denoted by $l$, the distribution function of the
kinetic energy at a total energy $E$, in
the harmonic approximation, is given by \cite{Labas,Pearson}:
\begin{equation}
g_{E,l}(K) = C_{l} (E-\Delta _{l}-K)^{\frac{3N-7}{2}} 
K^{\frac{3N-7}{2}}, 
\label{ghar}
\end{equation}
where  $\Delta _{l}$ is the potential energy of the isomer
$l$, relative to the potential energy value of the 
lowest-energy isomer, and $C_l$ is a normalization constant
such that:
\begin{equation}
\int_ {0}^{E- \Delta _{l}} g_{E,l}(K) dK = 1. 
\label{Cghar}
\end{equation}
The distribution function of the kinetic energy $g_{E,har}$, at a total
energy $E$, corresponding to the whole PEL can be constructed
by adding up the contribution of each harmonic potential well,
weighted by the probability, $\omega _{E,l}$,
of finding a given isomer during
the quenching from the MD trajectories. This probability
is given by the height of the distribution shown in Fig. 5.
Then, $g_{E,har}$ is given by:
\begin{equation}
g_{E,har} = \sum _{l=1}^{l_{max}} \omega _{E,l} \, g_{E,l}(K), 
\label{gharE}
\end{equation}
with
\begin{equation}
\sum _{l=1}^{l_{max}} \omega _{E,l} = 1.  
\label{wharE}
\end{equation}
By using the data from the whole distribution of isomers in Fig. 5,
the distribution function of the kinetic energy 
$g_{E,har}$ was calculated
for the Cd$_{55}$ and Na$_{55}$ clusters. They are displayed in the
insets of Fig. 5 (full lines). 
A comparison between $g_{E,har}$ and the exact
$g_E$ (obtained from the MD simulation and displayed
in the lower right inset of Fig. 4) shows that there is
a good agreement between the two distribution functions.
This indicates that $g_E$ is determined mainly from the
number of isomers and the probability to found them
(height of the distribution), 
rather than from other features of the PEL like saddle points.
The advantage in introducing $g_{E,har}$ in this discussion is
related with the fact that it is possible to analyze the
broadening of this distribution function of the kinetic energy
by considering different
subsets of potential energy minima (isomers). This is
useful to determine
what regions of the PEL are more relevant to increase the
width of $g_{E,har}$, and investigate the appearance of the
negative heat capacity.
The insets of Fig. 5 show three partial distribution functions
$g_{E,har}$, considering different subsets of isomers 
corresponding to three intervals of low (L), medium (M)
and high (H) potential energy values.
By analyzing the relative contribution of these subsets to the
width of $g_{E,har}$ for the Cd$_{55}$ and Na$_{55}$ clusters,
it is found that the larger broadening in the cadmium cluster is
mainly due to the larger contribution of the isomers in the range of
high potential energy which are spreaded along a larger
interval than those
corresponding to the Na$_{55}$ cluster. 
The width of $g_{E,har}$ for the Na$_{55}$
cluster is smaller since there are proportionally less isomers 
with high potential energy, and they are extended over a shorter
interval of values. 
As was mentioned above, the physical reason for this difference
in the distribution of isomers between the Cd$_{55}$ and Na$_{55}$
clusters is the shorter range of the many-body forces existing
in the cadmium cluster as compared with those present in the
sodium cluster. A similar result was obtained for 55-atom
clusters using a pairwise Morse potential for different values
of the range of the interatomic forces \cite{Doye}.
In that case the
backbending loop in the caloric curve (negative heat
capacity) was obtained using a Morse potential with a range
of the interatomic forces that is shorter than the range
characteristic of alkali metals which have long-ranged
interactions \cite{Doye}.

Therefore, if a detailed characterization of the distribution
of isomers forming the PEL of clusters is performed, 
the broadening of $g_{E}$ 
may be estimated, and by the comparison of the corresponding
$Z_{E}^{(2)}$ and $Z_c$ values, it would be possible to
predict the sign of
the heat capacity of clusters.

\section{Summary}
The microcanonical heat capacity of sodium clusters has been
calculated using constant-energy MD simulations and 
the many-body Gupta
potential. Negative values for the heat capacity at
energies where the cluster is at the melting-like transition
were found for Na$_{135}$, Na$_{142}$, and Na$_{147}$. The
smaller sodium clusters Na$_{N}$, $N$=13, 20, and 55, do not
show this peculiarity. An analysis of the calculated distribution
function of the kinetic energy $g_E$ for the six clusters investigated,
shows that the width of this distribution function 
is a useful
feature to determine the sign of
the heat capacity. It was found, that although the broadening of
$g_E$ is larger for the smaller clusters, it is not enough to
overcome  the corresponding threshold value
to obtain a negative heat capacity. However, since this threshold
is a fast decreasing function of the cluster size, the broadening
of $g_E$ in the larger clusters is enough to generate
a negative heat capacity.

It was also shown that the broadening of $g_E$ depends on the
distribution of potential energy minima that characterize the
PEL of clusters. Specifically, as the range of the many-body interactions
is decreased (like in the case of Cd clusters), the number
of local minima with higher energy increases generating a larger
broadening in $g_E$, and consequently a negative heat capacity.
The analysis presented in this paper shows 
how the complex topology of the PEL can be
explored to extract the main features that determine the
sign of the heat capacity of metal clusters.

\begin{acknowledgments}

This work was supported by Conacyt-Mexico under
Project No. G32723-E. JARN acknowledges
a graduate fellowship from DGEP-UNAM.

\end{acknowledgments}

\newpage

\begin{figure}
\caption{ \label{Fig1} Heat capacity of Na$_{N}$, $N$ = 13 (a); 20 (b);
55 (c); 135 (d); 142 (e); and 147 (f) clusters. 
The cluster energy is calculated taking
as reference the value of the binding energy of the most-stable
(lowest-energy)
configuration given in Table I of Ref. \onlinecite{Reyes}. } 
\end{figure} 

\begin{figure}
\caption{ \label{Fig2} Energy dependence of the $Z_E$ (black dots)
and $Z_{E}^{(2)}$ (stars) values for
Na$_{N}$, $N$ = 13 (a); 20 (b);
55 (c); 135 (d); 142 (e); and 147 (f) clusters.
The $Z_E$ values were calculated using Eq. (5) whereas the
$Z_{E}^{(2)}$ values, which are an approximation
of $Z_E$ according to Eq. (9), were obtained using
the second moment of the distribution function
of the kinetic energy. See the related text for an explanation
of the difference between these quantities.
The cluster energy is calculated taking
as reference the value of the binding energy of the most-stable
(lowest-energy)
configuration given in Table I of Ref. \onlinecite{Reyes}. } 
\end{figure} 

\begin{figure} 
\caption{ \label{Fig3} Distribution function of the kinetic energy
for Na$_{N}$, $N$ = 13 (a); 20 (b);
55 (c); 135 (d); 142 (e); and 147 (f) clusters.
The three curves displayed in each panel correspond to the
solid- (lower energy), melting- (intermediate energy), and
liquid-like (higher energy) phases. } 
\end{figure} 

\begin{figure}
\caption{\label{Fig4} Comparison of the maximum values of
$Z_{E}^{(2)}$ (black dots for the Na$_N$ clusters) and the threshold
$Z_c$ (continuous line) as a
function of the cluster size.  
The star shows the maximum value of
$Z_{E}^{(2)}$ for the Cd$_{55}$ cluster. 
The upper insets show the energy dependence of the
caloric curve and the heat capacity of the Cd$_{55}$
cluster. The inset at the low right corner shows the
$g_E$ as a function of the normalized mean deviation of
the kinetic energy of the Cd$_{55}$ cluster, and its comparison
with Na$_{55}$, calculated at E = 4.37 eV and E = 2.71 eV,
respectively.} 
 
\end{figure}

\begin{figure}
\caption{ \label{Fig5} Normalized distribution of 
potential energy minima
for the Na$_{55}$
and Cd$_{55}$ clusters. 
This distribution was obtained from two thousand
quenchings separated by 10000 time steps during a MD
trajectory of 20  million time steps.
The values of E=4.37 eV (Cd$_{55}$) and
E=2.71 eV (Na$_{55}$) correspond to the cluster energies when they
are at the middle of the melting-like transition.
The vertical dashed lines separate intervals of
low (L), medium (M), and high (H) potential energy.
The insets show the total (full line), including the
L, M, and H intervals, and partial (taking different
subsets of isomers) $g_{E,har}$, as a function of
the normalized mean kinetic energy.} 
\end{figure}

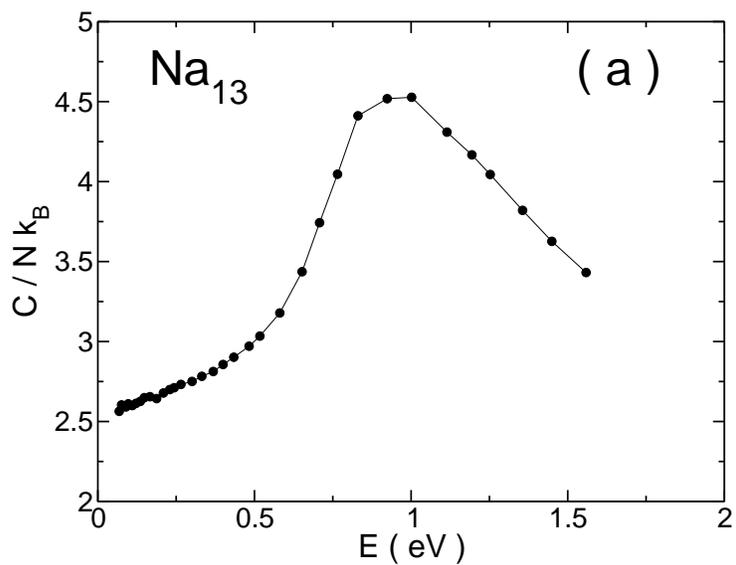
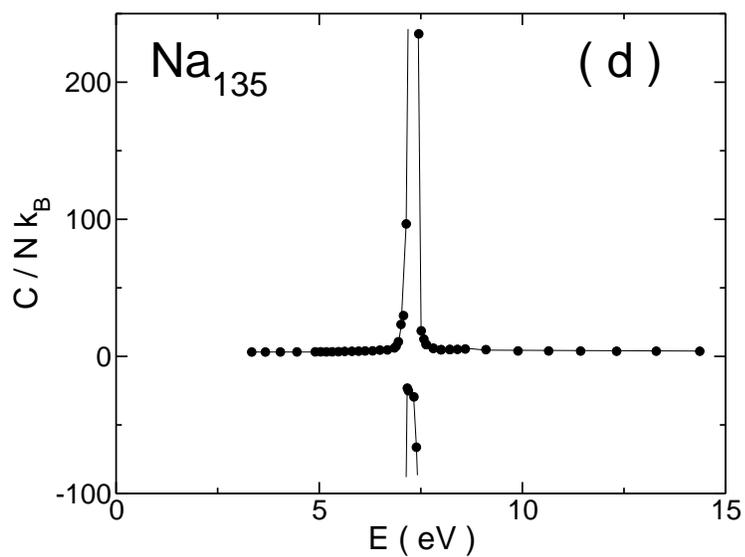
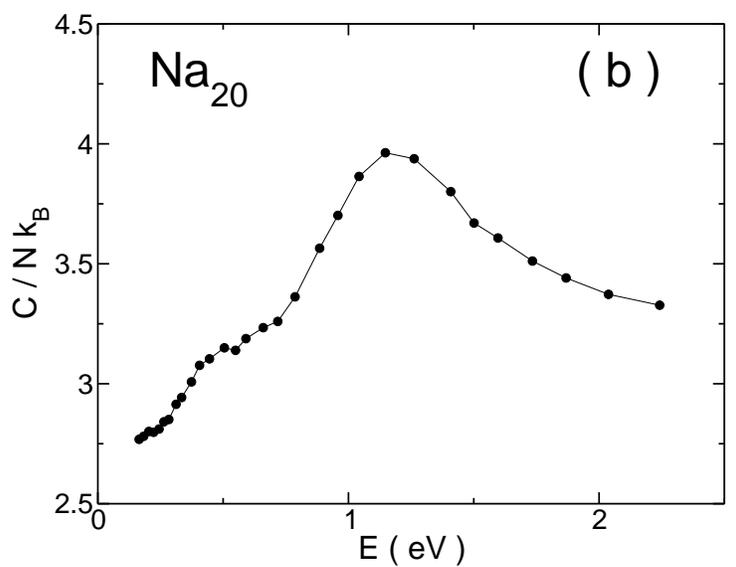
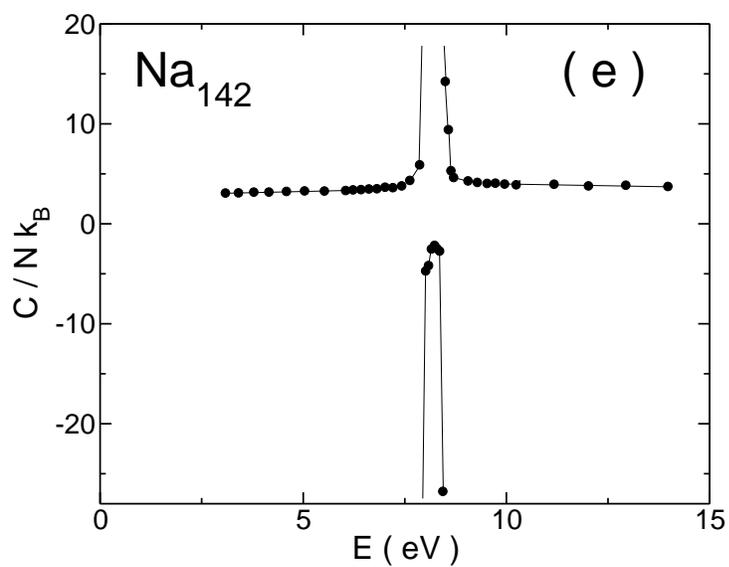
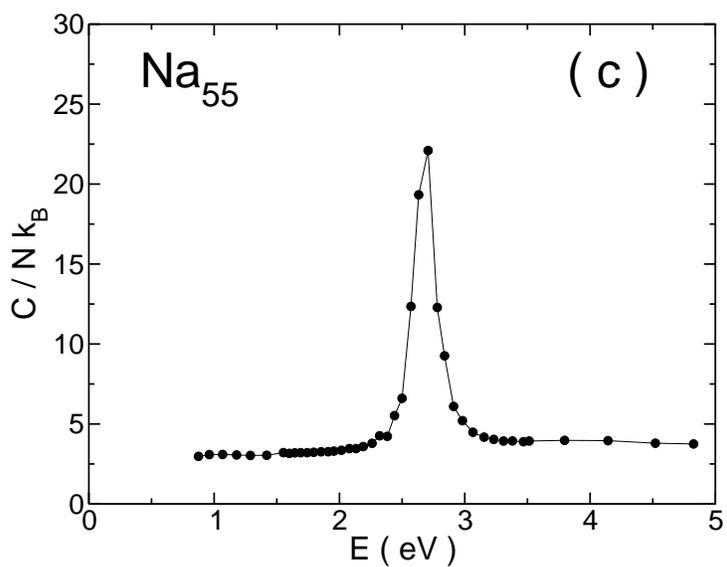
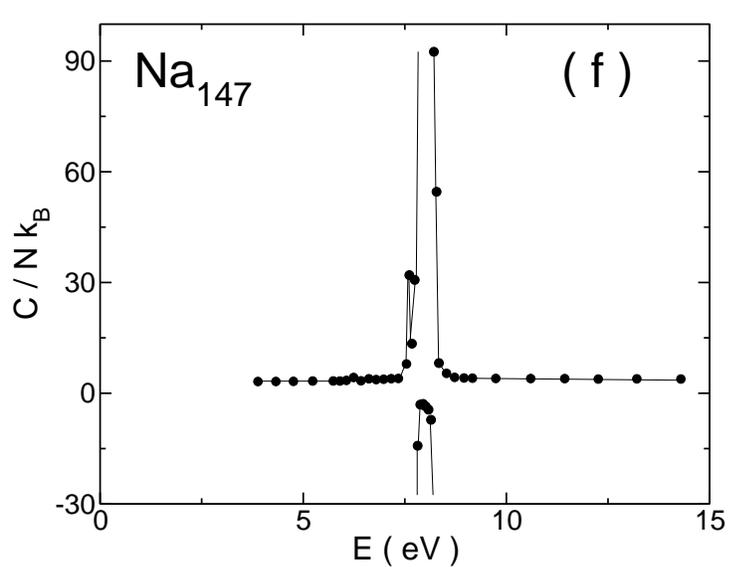
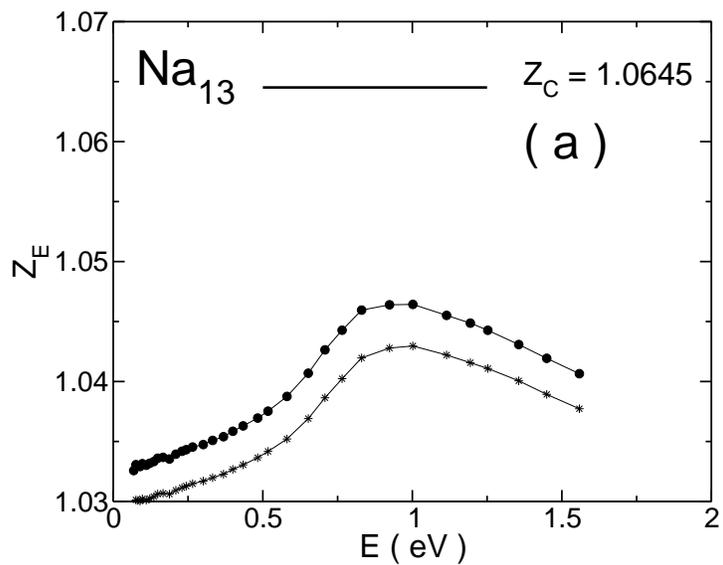
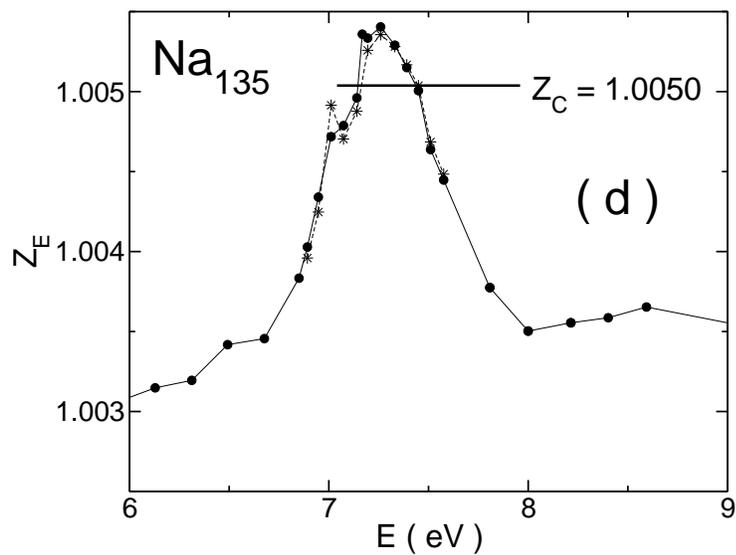
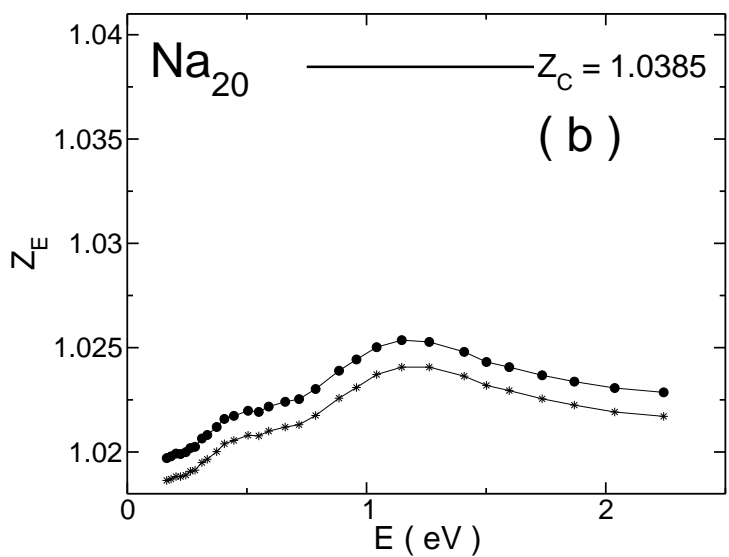
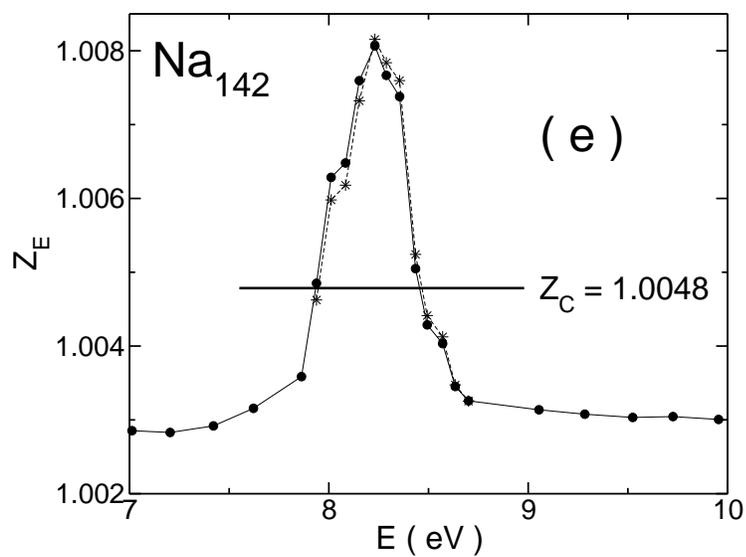
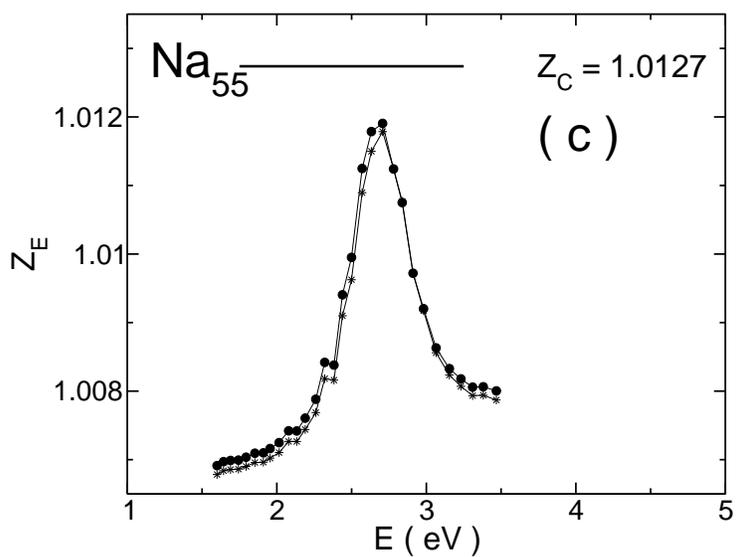
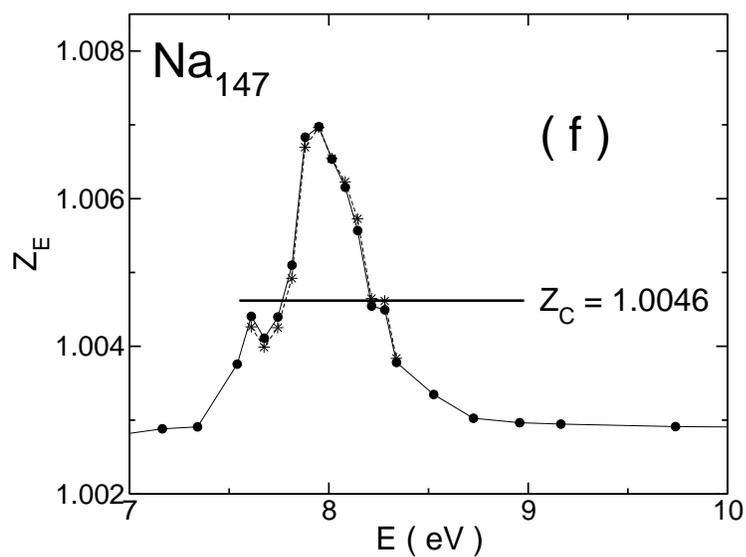
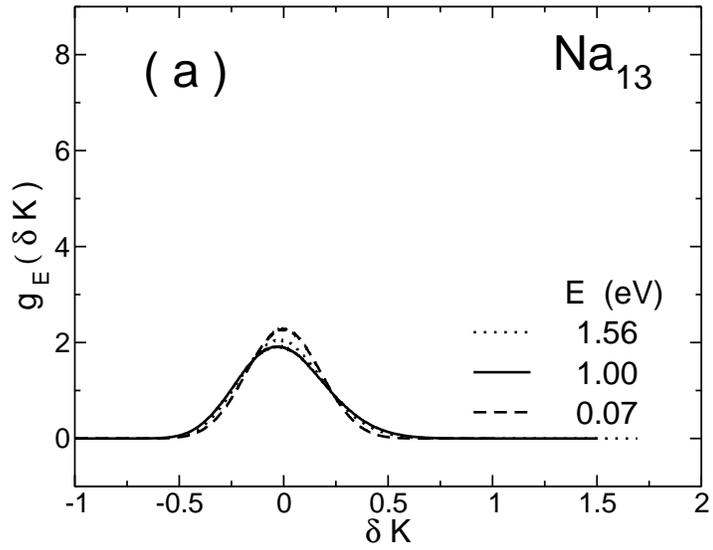
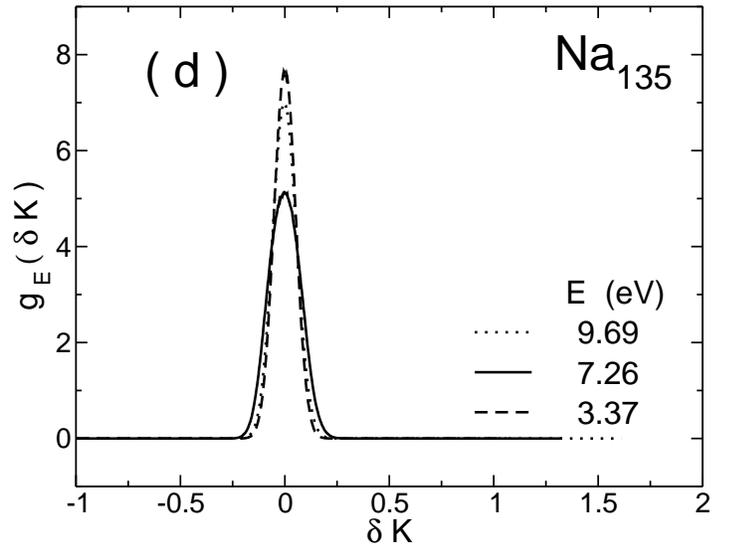
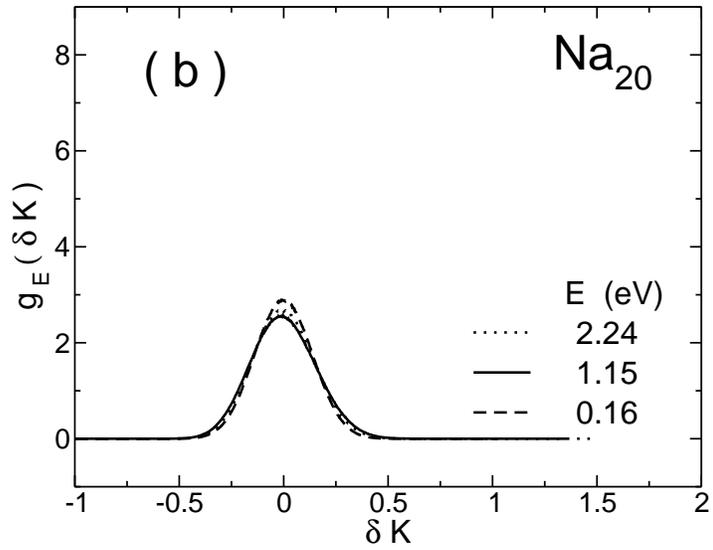
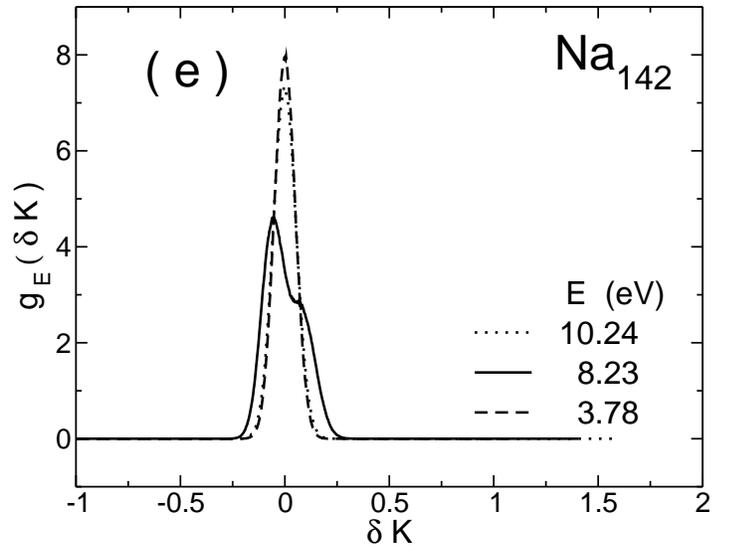
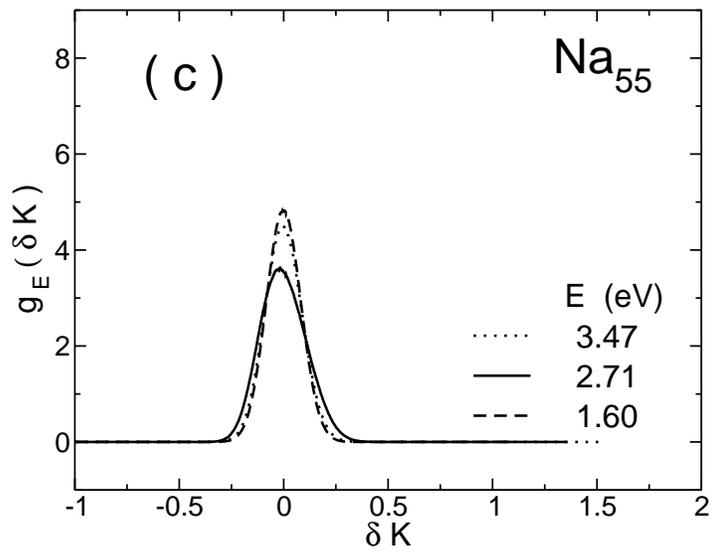
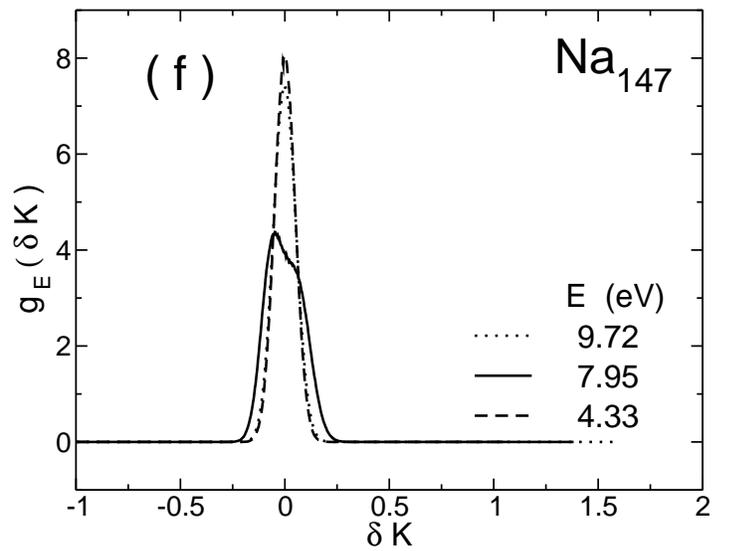
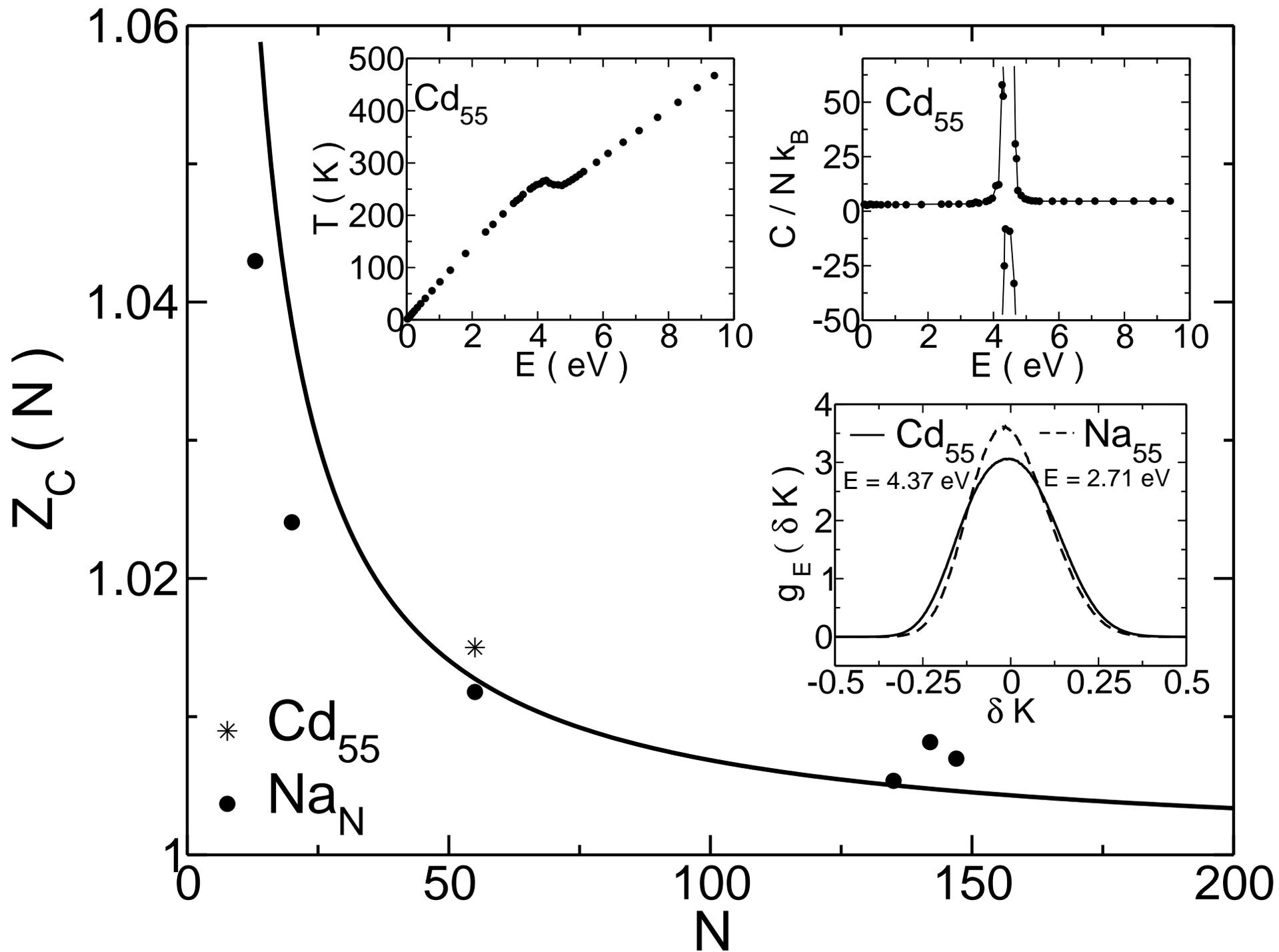
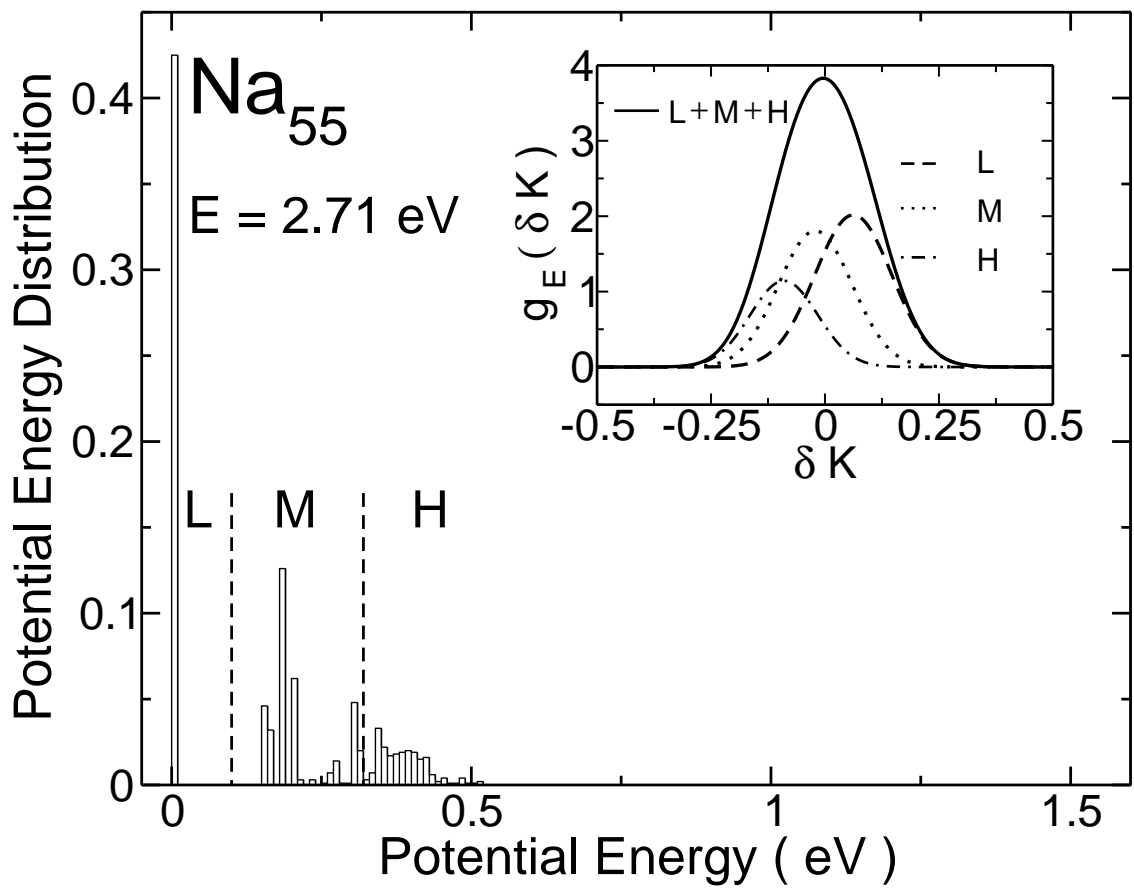
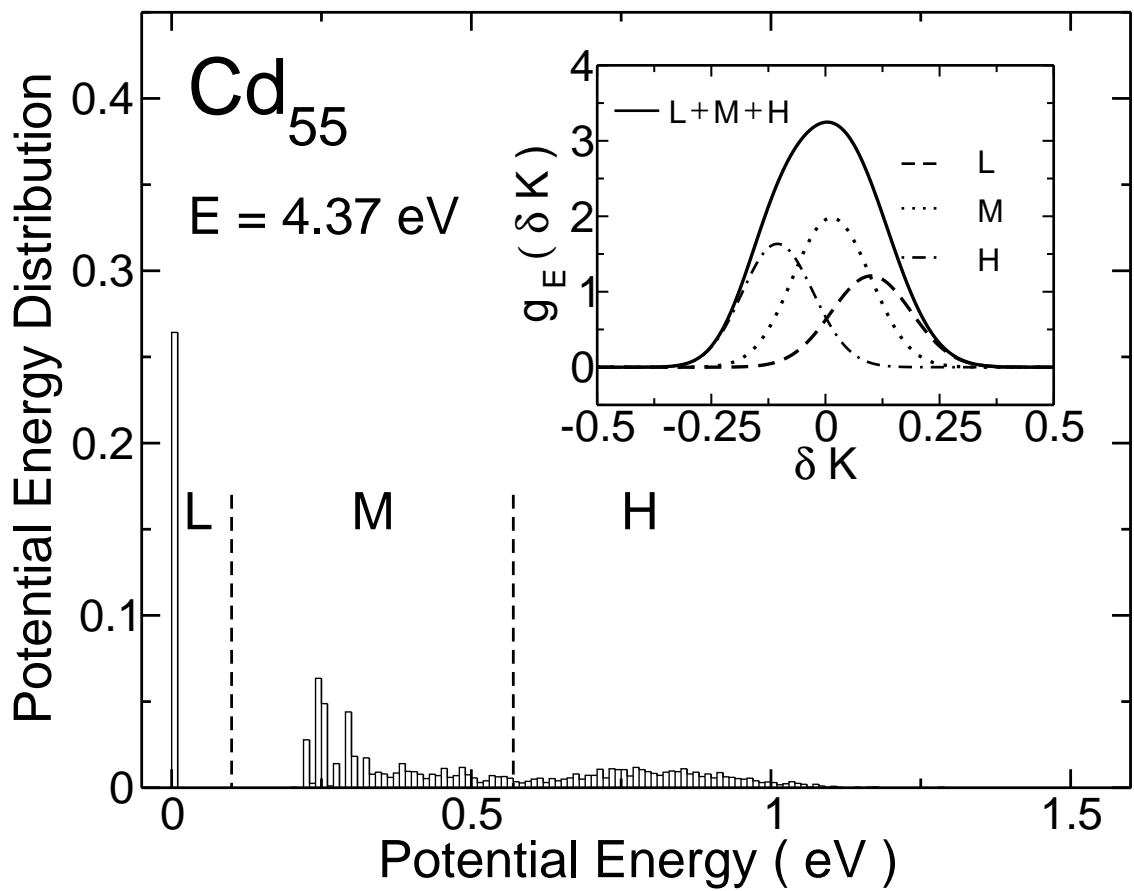
\end{document}